# Localization-delocalization transition in a one-dimensional system with long-range correlated off-diagonal disorder

H. Cheraghchi,[1,2],* S. M. Fazeli,[1] and K. Esfarjani[1]

[1]*Department of Physics, Sharif University of Technology, P.O.Box 11365-9161, Tehran, Iran*
[2]*Department of Physics, Damghan University of Basic Sciences, Damghan, Iran*

The localization behavior of the one-dimensional Anderson model with correlated and uncorrelated purely off-diagonal disorder is studied. Using the transfer matrix method, we derive an analytical expression for the localization length at the band center in terms of the pair correlation function. It is proved that for long-range correlated hopping disorder, a localization-delocalization transition occurs at the critical Hurst exponent $H_c = 1/2$ when the variance of the logarithm of hopping "$\sigma_{\ln(t)}$" is kept fixed with system size $N$. Numerically, this transition can be expanded to the vicinity of the band center. Based on numerical calculations, finite-size scaling relations are postulated for the localization length near the band center ($E \neq 0$) in terms of the system parameters $E, N, H$, and $\sigma_{\ln(t)}$.

## I. INTRODUCTION

According to the pioneering work of Anderson[1] in 1958, all electron states in one dimension (1D) are exponentially localized for any amount of uncorrelated diagonal disorder.[2] This expression can be also applied to off-diagonal disorder.[3] However, an additional sublattice symmetry around of the band center, which is called chiral symmetry, is the reason that one can separate the localization properties of purely off-diagonal disorder from those of the standard Anderson model.[4–6] This symmetry causes peculiar properties such as divergence of the density of states[7] and the localization length[8] at the band center.

Recently, interest in 1D disorder models with correlated disorder has been growing, as it has become progressively clear that correlations of the random potential can deeply affect localization properties. Spatial correlation of disorder can unexpectedly create extended states at some particular energies. In a system with short-range correlation of the on-site disorder—e.g., in the random dimer model[9]—one can have a discrete number of extended states. This demonstration attracted a great attention to investigate the existance of metal-insulator transitions in 1D disordered systems. Experimental evidence of the delocalization effect produced by these short-range correlations was recently found in semiconductor superlattices.[10]

Special attention has been recently paid to the presence of a continuum of extended states and mobility edges in the *long-range* correlated disorder model. In Ref. 11 de Moura and Lyra considered the discrete Anderson model with a self-affine potential as given by considering the potential as the trace of a fractional Brownian motion and imposing a normalization condition that kept fixed the variance of potentials for all system sizes. They showed how long-range correlated sequences of site energies could result in a continuum of extended states (although these results caused some controversy[12]). Furthermore, they numerically showed[13] that delocalization in the presence of pure long-range correlated hopping disorder requires weaker correlations than diagonal disorder.

Several stochastic processes in nature are known to generate long-range correlated random sequences which have no characteristic scale—for example, in the nucleotide sequence of DNA molecules[14] and height-height correlations in the interface roughness appearing during growth.[15]

To understand the effect of long-range correlations of the disorder on a phase transition, Izrailev *et al.*[16] perturbatively derived an analytical relationship between the localization length and pair correlator of weak disorder. They showed how specific long-range disorder correlations lead to the appearance of mobility edges in 1D discrete models. An experimental confirmation of these findings was obtained by studying the transmission of microwaves in a single-mode waveguide with a random array of correlated scatterers.[17] By this method, Izrailev *et al.*[18] have also studied that the correlation on the positions of the scatterers, which correspond to off-diagonal disorder, induces mobility edges in a nonperiodic Kronig-Penny model for weak disorder.

Analytical results of the work of Izrailev *et al.*[16] and those which have followed[18–20] were obtained to second order in the disorder. Taking into account high-order terms, however, makes extended states disappear.[20] Furthermore, the perturbative approach fails close to the center and edges of the band.[16] Since the conductance of the system depends on the localization properties near the band center (Fermi level), an unperturbative analytical calculation at the Fermi energy is needed in order to investigate the true nature of the metal-insulator transition.

Another reason for the revival of interest in the 1D disordered model is due to the revision of the well-known single-parameter scaling (SPS) hypothesis.[21] According to this hypothesis, in the thermodynamic limit, the mean Lyapunov exponent (LE) and its variance become related. However, as has been shown in Ref. 22, the SPS hypothesis is violated in finite-length LE systems. Violation of the SPS hypothesis also occurs in the off-diagonal disorder case at the band center and near to it.[23,6]

In the present work, we have investigated the 1D Anderson model with correlated and uncorrelated hopping disorder analytically and unperturbatively at $E=0$ and numerically at other energy values near the band center. It will be proved that the probability distribution function of the LE at the band center is as a semi-Gaussian function in the uncorrelated case. According to this distribution function, we find that regardless of the probability distribution of hopping terms, in the case of uncorrelated disorder, the $E=0$ state is anomalously localized ($\lambda_{\text{localization}} \propto N^{1/2}$). This is in agreement with the work of Ref. 4.

In the correlated disordered case at zero energy, we have also derived an analytical expression for the localization length in terms of the pair correlation function. In the case of a log-normal distribution function of hopping disorder with *normalized* long-range correlations, we have proved that there is a localization-delocalization transition which has the same critical exponent as diagonal disorder $H_c=1/2$ for all energies near the band center (numerically in Fig. 7 and analytically at the band center). This is in contrast with the result of the authors of Ref. 13 which concluded that $H_c=0$. They have done no size scaling study, and therefore a conclusion in the thermodynamic limit is incorrect. Here $H_c$ is the critical value of the Hurst exponent defined later in Eq. (16).

The localization length is shown to have a power-law behavior versus size at the band center which has chiral symmetry. This power-law behavior also appears near the band center when states are delocalized for $H \geq 1/2$. Close to the band center and for $H < 1/2$, three regions are clear which are sorted by the length of the system (or- equivalently by energy or variance of disorder). We will show that the localization properties at $E=0$ can be extended to other energies around the band center if the size is smaller than a critical length, while systems with a size larger than the critical length are in the localized regime. In the localized regime, the divergence of the localization length versus energy near the band center has a simple power-law behavior with exponent $\eta=0.18\pm0.03$ for both correlated and uncorrelated disorder. The region between these regimes is called as crossover region.

This article is organized as follows: Section II focuses on our model and the definition of the Lyapunov exponent. Section III is an exact solution of hopping disorder at the band center for uncorrelated and correlated disorder. Finally, Sec. IV describes the numerical support for our analytical results and scaling properties of states close to the band center. The dependence of the localization length in terms of energy and variance near the band center is also investigated. We also expand the transition to other energies near the band center in this section. Discussions and conclusions are finally presented in Sec. V.

## II. MODEL AND DEFINITION OF THE LYAPUNOV EXPONENT

We consider noninteracting electrons in 1D disordered systems within a nearest-neighbor tight-binding formalism. The Schrödinger equation projected on site $i$ becomes

$$\varepsilon_i \psi_i + t_{i,i+1} \psi_{i+1} + t_{i-1,i} \psi_{i-1} = E \psi_i, \quad (1)$$

where $E$ is the energy of the incoming electron. $|\psi_i|^2$ is the probability of finding an electron at site $i$, $\varepsilon_i$ is the potential at site $i$, and $t_{i-1,i}=t_{i,i-1}$ is the hopping integral from site $i-1$ to site $i$. In the transfer matrix method, the above equation can be written in the recursive matrix form

$$\begin{pmatrix} \psi_{n+1} \\ \psi_n \end{pmatrix} = \begin{pmatrix} \frac{E-\varepsilon_n}{t_{n+1,n}} & \frac{-t_{n-1,n}}{t_{n+1,n}} \\ 1 & 0 \end{pmatrix} \begin{pmatrix} \psi_n \\ \psi_{n-1} \end{pmatrix}. \quad (2)$$

The wave functions of the two ends can be related together by calculating the product matrix as $T_{N,0}=\Pi_{i=1}^{N} T_{i,i-1}$, where $N$ is the sample length and $T_{i,i-1}$ is a transfer matrix which connects the wave functions of sites $i$ to $i-1$. One of the eigenvalues of $T_{N,0}$ ($a_{max}$) grows with the system size and the other one ($a_{min}$) decreases.[24] The Lyapunov exponent is defined by the norm of $T$ as[25]

$$\gamma = \lim_{N \to \infty} \frac{1}{2N} \langle \ln \| T_{N,0}^{\dagger} T_{N,0} \| \rangle_{c.a.}. \quad (3)$$

Here $\langle \cdots \rangle_{c.a.}$ refers to the configuration average. Hereafter, the label of hopping terms changes as $t_i=t_{i,i-1}$. It should be pointed out that the determinant of every transfer matrix is not unity, $\text{Det}[T_{n,n-1}]=t_n/t_{n+1}$. It is clear that the determinant of the total transfer matrix becomes $\text{Det}[T_{N,0}]=\Pi_{i=1}^{N}\text{Det}[T_{i,i-1}]=t_1/t_{N+1}$. Since $t_1$ and $t_{N+1}$ come from the same distribution function of disorder, without applying any boundary conditions, we can conclude that $\langle \ln\{\text{Det}[T_{N,0}]\} \rangle_{c.a.}=0$.

Therefore, the largest and smallest eigenvalues have asymptotically the following form:

$$\langle \ln|a_{max}| \rangle_{c.a.} = - \langle \ln|a_{min}| \rangle_{c.a.}. \quad (4)$$

The norm of $T$ [Eqs. (3) and (4)] is dominated by the largest eigenvalue.

$$\gamma = \lim_{N \to \infty} \frac{1}{N} \langle \ln(|a_{max}|) \rangle_{c.a.} = - \lim_{N \to \infty} \frac{1}{N} \langle \ln(|a_{min}|) \rangle_{c.a.}. \quad (5)$$

Without losing any generality, we can impose the boundary condition (BC) $t_1=t_{N+1}$ in order to have a unity determinant [Eq. (2)] for any configuration of disorder. This BC is also realized if the disordered system is attached to two perfect leads. If the hopping energy of both leads is taken to be unity, the above BC is automatically satisfied. Using this BC, the eigenvalues of the total transfer matrix are the inverse of each other, $a_{max}=1/a_{min}$, for any configuration of disorder. As will be shown later, we perform the configuration average on one of the eigenvalues of the total transfer matrix. It can either be $a_{max}$ or $a_{min}$ in every configuration. Consequently, it is clear that in order to calculate the Lyapunov exponent we have to average absolute values of the logarithm of any of the eigenvalues. Therefore, the final form for $\gamma$ becomes

$$\gamma = \lim_{N \to \infty} \frac{1}{N} \langle |\mathbf{F}| \rangle_{c.a.}, \quad (6)$$

where $\mathbf{F}=\ln(|\mathbf{a}|)$ and $\mathbf{a}$ is an arbitrary eigenvalue of the transfer matrix. In our model, we consider all on-site energies to be zero. The total transfer matrix can be easily derived at $E=0$ as follows:

$$T_{2k,0} = (-1)^k \begin{pmatrix} \frac{t_2 t_4 \cdots t_{2k}}{t_3 t_5 \cdots t_{2k+1}} & 0 \\ 0 & \frac{t_1 t_3 \cdots t_{2k-1}}{t_2 t_4 \cdots t_{2k}} \end{pmatrix}. \quad (7)$$

Here, the transfer matrix has been computed for even $N$ or an odd number of atoms.[26] The odd parity in number of atoms results in an eigenstate at $E=0$, while an even number of atoms does not have any eigenvalue at the band center. It can be proved by the symmetry property of the density of states (DOS) near the band center. Under a mapping $\psi_n \to (-1)^n \psi_n$, the Hamiltonian eigenvalue in Eq. (1) with zero on-site energies changes sign $E \to -E$. Consequently, the eigenvalues of the energy always occur in pairs around of the band center as $\pm E$. When the number of sites is odd, $E=0$ is one of the eigenvalues. This property is referred to the chiral (sublattice) symmetry. This symmetry follows when the lattice is divided into two sublattices $A$ and $B$ such that the hopping matrix only connects sites on sublattices, but not any connection in the same sublattice. The number of eigenstates at the band center is as $|N_A - N_B|$, where $N_A$ and $N_B$ are site numbers on each sublattices. It is clear that when $N = N_A + N_B$ is even such that $N_A = N_B$, there is no any eigenstate at the band center.[5]

Since we are interested in states at the band center, we will only consider wires with an odd number of atoms.[2] Therefore, the transfer matrix method is valid for an odd number of atoms at the band center. The function $F$ can be written as

$$\mathbf{F} = \ln\left(\frac{t_1 t_3 \cdots t_{2k-1}}{t_2 t_4 \cdots t_{2k}}\right) = \sum_{i=1}^{k} \left[\ln\left(\frac{t_{2i-1}}{t_0}\right) - \ln\left(\frac{t_{2i}}{t_0}\right)\right], \quad (8)$$

where $\ln(t_0) = \langle \ln(t_i) \rangle_{c.a.}$.

### III. EXACT SOLUTION OF HOPPING DISORDER AT $E=0$

#### A. Uncorrelated disorder

The central-limit theorem can be applied to uncorrelated disorder regardless of the distribution function of $t_n$. Hence, the sum in Eq. (8) has a Gaussian distribution function with zero mean:

$$P(F) = \frac{1}{\sqrt{2\pi\sigma_F^2}} \exp\left(-\frac{F^2}{2\sigma_F^2}\right), \quad (9)$$

where the variance of $\mathbf{F}$ is defined as

$$\sigma_F^2 = N \left\langle \left[\ln\left(\frac{t}{t_0}\right)\right]^2 \right\rangle_{c.a.} = N\sigma_{\ln(t)}^2. \quad (10)$$

By using Eq. (6), the Lyapunov exponent of such system is given by

$$\gamma = \sqrt{\frac{2}{\pi}} \frac{\sigma_F}{N}. \quad (11)$$

Therefore the localization length increases with system size as

$$\lambda = \frac{1}{\gamma} = \sqrt{\frac{\pi}{2}} \frac{N^{1/2}}{\sigma_{\ln(t)}}. \quad (12)$$

In the thermodynamic limit, this state is "anomalously" localized and $a_{max}$ has the exponential form $\exp(\xi N^{1/2})$. Actually, the localization length increases with system size, but it is always smaller than it. The localization length is also inversely proportional to $\sigma_{\ln(t)}$. Based on numerical simulations, we will show later that away from $E=0$ the localization length changes with variance as $1/\sigma_{\ln(t)}^2$. The authors in Ref. 8 have also obtained an equation similar to Eq. (8), but their expression of the LE refers to the mean value of the $\mathbf{F}$ function. Hence, they have wrongly concluded an extended state at zero energy.

#### B. Long-range correlated disorder

In the correlated case, the central-limit theorem fails and one needs to consider a special distribution function for $\mathbf{t}_n$. Of course, we are still able to use the result of Eq. (11) by taking a Gaussian distribution for $\ln(t)$'s:

$$P(\ln(t)) = \frac{1}{\sqrt{2\pi}\sigma_{\ln(t)}} \exp\left[-\frac{[\ln(t/t_0)]^2}{2\sigma_{\ln(t)}^2}\right]. \quad (13)$$

According to Wick's theorem, the distribution function of $\mathbf{F}$ will be Gaussian if every term in the summation has a Gaussian distribution. Therefore, the LE has again a semi-Gaussian probability distribution. Now, calculation of the localization length just requires one to calculate $\sigma_F$. The Lyapunov exponent can be derived in terms of the pair correlation function of $\ln(t)$ (See the Appendix for details):

$$\gamma = \frac{1}{N} \sqrt{\frac{2}{\pi}\left\{Ng(0) + 2\sum_{\ell=1}^{N-1} (N-\ell)(-1)^\ell g(\ell)\right\}},$$

$$g(i-j) = \left\langle \ln\left(\frac{t_i}{t_0}\right) \ln\left(\frac{t_j}{t_0}\right) \right\rangle_{c.a.}. \quad (14)$$

This equation is converted to the uncorrelated case [Eq. (12)] with $g(\ell) = 0$ for $\ell \neq 0$ and $g(0) = \sigma_{\ln(t)}^2$. It is necessary that we assume more special cases to determine the pair correlation function. From now on, long-range correlated disorder, which has been studied in Refs. 11, 13, and 27, will be considered.

The hopping terms are usually supposed to decay exponentially with bond length:

$$t(r_{nm}) = t_0 \exp(-\alpha r_{nm}). \quad (15)$$

Here $\alpha$ is the damping coefficient and $r_{nm}$ is the bond length between sites $n$ and $m$. So a Gaussian distribution for $r_{nm}$ results in a log-normal distribution for hopping amplitudes. Randomness is imposed on the $\ln t$'s which have a normal distribution. The rescaled self-affine landscapes for $\ln t$'s are given by the trace of a fractional Brownian particle with Hurst exponent $H$.[11] The fluctuations of the $\ln(t)$'s are given by

$$\left\langle \left[\ln\left(\frac{t_i}{t_0}\right) - \ln\left(\frac{t_j}{t_0}\right)\right]^2 \right\rangle_{c.a.} = 2\sigma_{\ln(t)}^2 \left|\frac{i-j}{\ell_c}\right|^{2H}, \quad (16)$$

where $\sigma_{\ln(t)}^2$ is kept fixed for all system sizes.[12] The correlation length ($\ell_C$) is considered to be equal to the system size. $i$ and $j$ are the positions of the bonds along the chain. The pair correlation function arising from Eq. (16) results in the following expression for Lyapunov exponent from Eq. (14):

$$\gamma = \frac{\sigma_{\ln(t)}}{N^{1/2}} \sqrt{\frac{4}{\pi} \sum_{\ell=1}^{N-1} (-1)^{\ell+1} \left(\frac{\ell}{N}\right)^{2H} \left(1 - \frac{\ell}{N}\right)}. \quad (17)$$

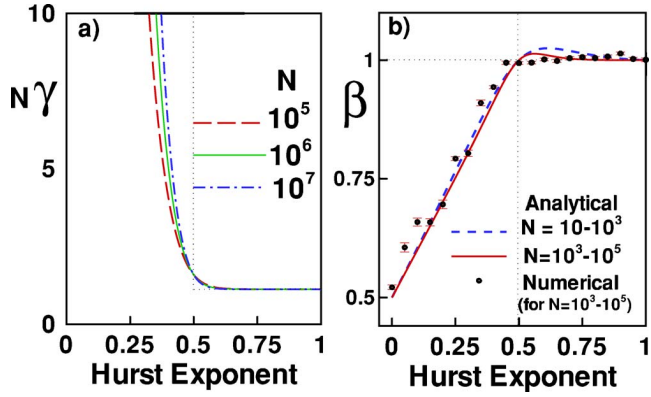

FIG. 1. (Color online) Localization-delocalization transition for long-range correlated disorder at $E=0$. (a) shows $N\gamma$ versus $H$ (Hurst exponent) for different sizes of the system (b) Exponent of $N$ in ($\lambda \propto N^\beta$) versus Hurst exponent. Finite-size effects near the critical Hurst exponent can be removed in the thermodynamic limit (analytical result).

This equation shows a sharp phase transition between delocalized-localized states at the critical exponent $H_{cr}=1/2$. Figure 1(a) shows $N\gamma$ versus $H$ (Hurst exponent) for different sizes of the system. For $H>1/2$, $N\gamma$ remains fixed with increasing system sizes, while for $H<1/2$, it increases with size. The situations $H>1/2$ and $H<1/2$ correspond to delocalized and localized states, respectively. The localization length has asymptotically a power-law behavior as $N^\beta$ for all $H$. $\beta$ equal to unity leads to extended states, while $\beta<1$ corresponds to localized states. Figure 1(b), which is plotted by using Eq. (17), is related to the variation of $\beta$ in different regions of $H$. The finite-size effect near the critical exponent ($H_{cr}$), which is shown in Fig. 1(b), can be removed in the thermodynamic limit. Hence, in the thermodynamic limit, the localization length is proportional to $N$ for $H>1/2$ and $N^\beta$ for $H<1/2$ where $\beta=\frac{1}{2}+H<1$:

$$\lambda \propto \begin{cases} \frac{N}{\sigma_{\ln(t)}}, & H > \frac{1}{2}, \\ \frac{N^{1/2+H}}{\sigma_{\ln(t)}}, & H < \frac{1}{2}. \end{cases} \quad (18)$$

Therefore, the localization length has a smooth variation in exponent from the localized regime ($H<1/2$) to the strongly correlated regime ($H>1/2$). The recently mentioned results can be summarized by the following formula for the localization length:

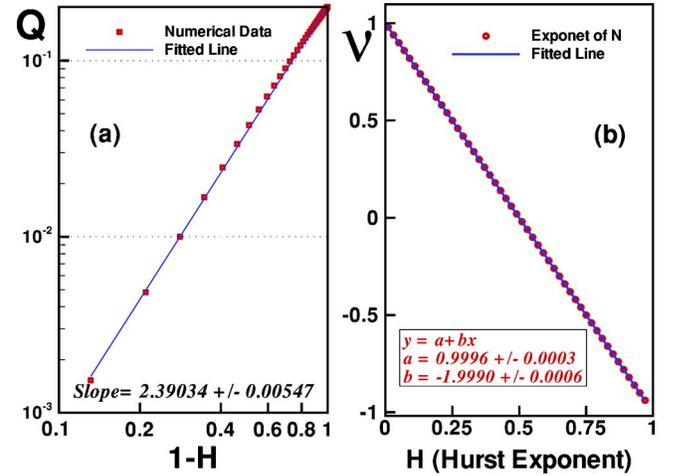

FIG. 2. (Color online) (a) Exponent of $N$ and (b) its coefficient in the relation $[(N\gamma/\sigma_{\ln(t)})^2 - 1/\pi] \propto Q N^\nu$.

$$\lambda \propto \frac{1}{\sigma_{\ln(t)}} \frac{N}{\sqrt{\frac{1}{\pi} + Q(H) \times N^{1-2H}}}. \quad (19)$$

This equation fairly describes the behavior of the curve in Fig. 1(b). The above equation is extracted from Eq. (17) when one investigates the following proportionally:

$$\left[\left(\frac{N\gamma}{\sigma_{\ln(t)}}\right)^2 - \frac{1}{\pi}\right] \propto Q(H) N^{\nu(H)}. \quad (20)$$

In the $H>1/2$ region and in the thermodynamic limit, the quantity of $(N\gamma/\sigma_{\ln(t)})^2$ tends to the value of $1/\pi$. In order to derive $Q$ and $\nu$, it requires one to find the slopes and intercept points of the lines which are defined by a linear relation between $\ln[(N\gamma/\sigma_{\ln(t)})^2 - 1/\pi]$ and $\ln(N)$. Figure 2 shows the slopes $[\nu(H)]$ and intercepts $[\ln(Q)]$ of lines versus $H$. The fitted line in Fig. 2(a) shows that the dependence of $\nu$ is as $\nu(H)=1-2H$. The coefficient ($Q$) of $N$ in the above equation is also plotted in Fig. 2(b) in terms of $H$ ($<1$). As the fitted line shows, the coefficient in the above equation follows from the form of $Q(H) \propto (1-H)^\chi$, where $\chi = 2.39034 \pm 0.00547$.

It should be mentioned that $\sigma_{\ln(t)}$ in Eq. (17) or (19) does not affect localization properties; nor does it induce any phase transition unlike the claim of Shima *et al.* in the case of on-site disorder.[28] It is only a coefficient independent of $N$ in the localization length, but will, however, affect the transmission coefficient. It can be concluded from Eq. (18) that, if we impose no constraint on the variance $\sigma_{\ln(t)} \propto N^H$,[12,29] all states become localized for any value of $H$. In fact, the localization length behaves as $N^{1/2}$ for all $H$ in the region below the critical exponent ($H<1/2$). The state is more localized ($\lambda \propto N^{1-H}$) in the region above the critical exponent ($H>1/2$).

## IV. NUMERICAL RESULTS

The Lyapunov exponent as defined in Eq. (3) has been calculated numerically in a model where hopping amplitudes were defined by Eq. (15). In this section, we will check our analytical results at the band center by numerical calculations. Also, our studies, such as the localization-delocalization transition, could be numerically extended to energies close to the band center.

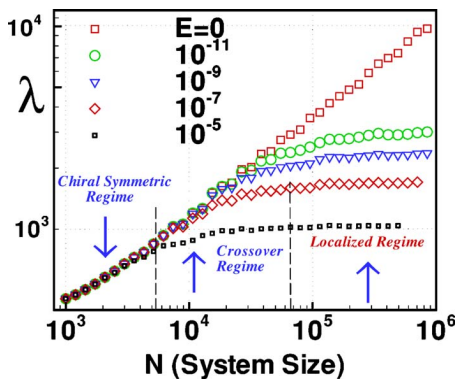

FIG. 3. (Color online) Localization length as a function of system size for the uncorrelated case for several energy values. The variance of $\sigma_{\ln(t)}$ is fixed to be 0.1.

As was mentioned in the Introduction, there exists an addition symmetry at the band center which is called the chiral (sublattice) symmetry. At energies close to the band center, chiral symmetry is broken. At a sufficiently long lengths of the system, its localization properties will flow to those of the standard symmetry class. This flow is governed by a critical length scale ($N_{cr}$) so that for $N \ll N_{cr}$ the localization properties are still like those of the chiral symmetry class (band center behavior), while for $N \gg N_{cr}$, the localization properties are determined by the standard symmetry class (localized regime). In summary, we distinguish three different regimes[6]: (1) the region with a chiral symmetry in $N \ll N_{cr}$, (2) the crossover region in a length $N \approx N_{cr}$ is the boundary which separates the regime with nonstandard symmetry from the standard symmetry regime, and (3) the localized regime, which follows the standard symmetry class, will occur in $N \gg N_{cr}$. This critical length is equivalent to a critical energy and variance. In the following sections, we find the scaling properties of these regions for both uncorrelated and correlated disorder.

### A. Uncorrelated disorder

In this section, we study the property of the localization length versus system size, energy, and variance. The scaling relations can be derived from the localization properties of the band center (which was studied in analytical calculations) for $N \ll N_{cr}$ and from the numerical calculations in the region $N \gg N_{cr}$.

Figure 3 shows $\lambda(N)$ for several values of the energy. It is clear that if we want to determine the localization behavior far from the band center, the system will exhibit a dual treatment as a function of system size. There is a critical length ($N_{cr}$) below which the localization length is proportional to $N^{1/2}$. In this region, the system behaves as if the energy were at the band center with chiral symmetry. Above the critical length, states show strongly localized behavior. On the other hand, a specified localization length, which is constant for all system sizes, is attributed to the system. The crossover region which separates the above two regions cannot be treated by our numerical calculations. The critical length has an inverse relation with the energy and strength of the disorder. At the band center, the critical length tends to infinity. Hence, the localization length is proportional to $N^{1/2}$ at $E=0$ for all system sizes. Therefore, the result in Eq. (12) is confirmed by the numerical evidence.

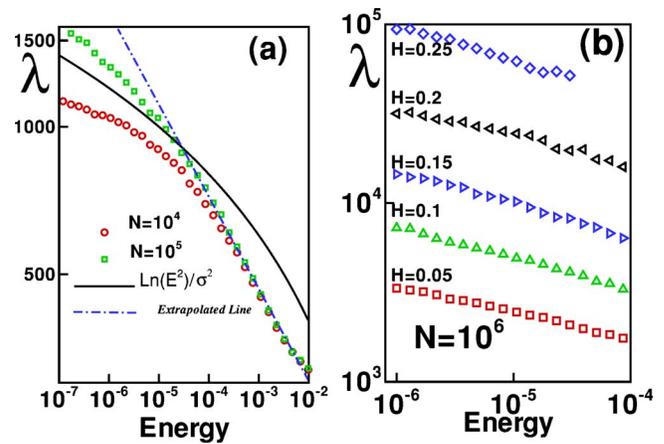

FIG. 4. (Color online) Localization length as a function of energy (near the band center) (a) in the case of uncorrelated disorder for several values of the system sizes and (b) in the case of correlated disorder for several values of the Hurst exponent $H$. $\sigma_{\ln(t)}$ is fixed to be 0.1.

The energy dependence of the function $\lambda(E)$ can also be evaluated for a certain system size. This system size should be large enough so that for all considered energies $E \gg E_{cr}$, one is in the localized regime. We define $E_{cr}$ by $N_{cr}(E_{cr})=N$. From Fig. 3 it seems that $N=10^5$ is greater than the critical length for all energies $E \gtrsim 10^{-5}$. Figure 4(a) suggests that the exponent ($\eta$) of the energy in the localized regime $\lambda(E) \propto E^{-\eta}$ is $\eta \approx 0.18 \pm 0.03$. In the thermodynamic limit, $\lambda(E)$ tends to form as an extrapolated line shown in Fig. 4(a). All points located on the extrapolated line are in the localized regime. A logarithmic divergence of the localization length versus energy was derived through the Thouless equation by Ref. 8 as the form of $2|\ln(E^2)|/\sigma^2$. This approximated logarithmic form, which has been derived in the thermodynamic limit, is not in good correspondence with the extrapolated line.

The functional of the localization length with respect to the variance $\lambda(\sigma_{\ln(t)})$ changes from the band center energy to other energies. As we proved before, the localization length is proportional to $\sigma_{\ln(t)}^{-1}$ at the band center. This functional is obvious in Fig. 5(a). Far from the band center, it has been reported[8] that the localization length is proportional to $\sigma_{\ln(t)}^{-2}$. Also, at energies very near the band center and for a fixed system size, the exponent changes from $(-1)$ for $\sigma \ll \sigma_{cr}$ to $(-2)$ for $\sigma \gg \sigma_{cr}$, where $\sigma_{cr} \propto E^{-\eta}/\sqrt{N}$. The lines in Fig. 5(a) break in the critical variance which for $\sigma \gg \sigma_{cr}$—one is in the localized regime, while the chiral symmetry regime will be observed in the region $\sigma \ll \sigma_{cr}$.

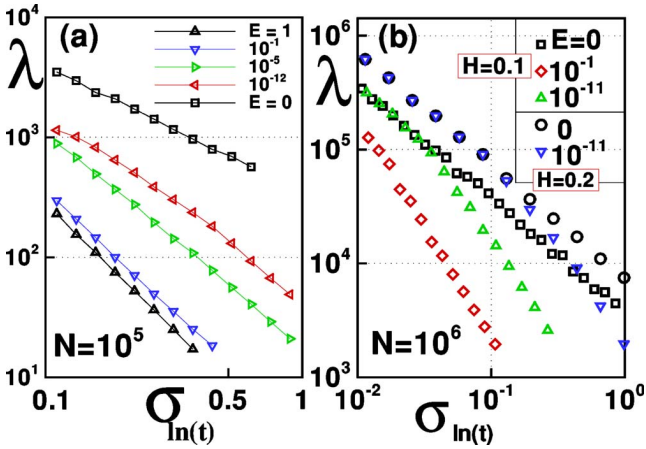

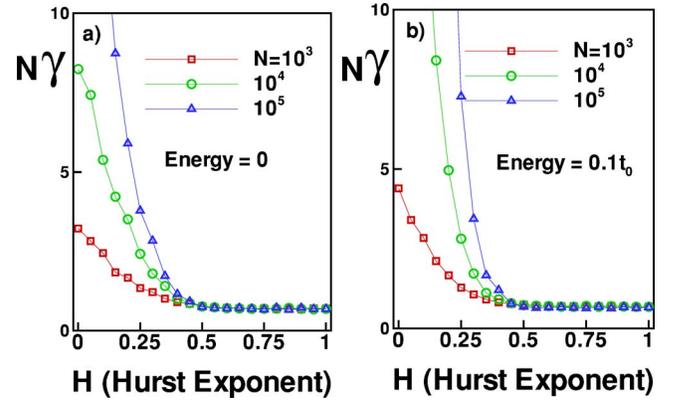

FIG. 5. (Color online) Localization length as a function of variance $\sigma_{\ln(t)}$ for several values of energy. (a) The case of uncorrelated disorder with fix system size $N=10^5$. (b) The case of correlated disorder for various of the Hurst exponent $H$.

We can summarize the above results in the following scaling law for $E \neq 0$:

$$\lambda \propto \begin{cases} \dfrac{N^{1/2}}{\sigma_{\ln(t)}}, & N \ll N_{cr}, \\ \dfrac{E^{-\eta}}{\sigma_{\ln(t)}^2}, & N \gg N_{cr}, \end{cases} \quad (21)$$

where $\eta \approx 0.18 \pm 0.03$. Consequently, the above equation results in the following form for $N_{cr}$:

$$N_{cr}(E,\sigma_{\ln(t)}) \propto \frac{E^{-2\eta}}{\sigma_{\ln(t)}^2}. \quad (22)$$

For $E=0$, as confirmed by both analytical and numerical evidence, $\lambda \propto N^{1/2}/\sigma_{\ln(t)}$ for all system sizes.

### B. Long-range correlated disorder

A sequence of long-range correlated variations of the bond length $\{r_i\}$ in Eq. (15) is produced by the Fourier filtering method.[30] This method is based on a transformation of the Fourier components ($\theta_k$) of a random number sequence $\{\theta_i\}$. The inverse Fourier transformation of the sequence $\{r_k\}(=k^{-(1/2+H)}\theta_k)$ leads to the sequence of interest $\{r_i\}$. The resulting variations of the bond lengths are spatially correlated with spectral density $S(k) \propto k^{-(1+2H)}$. It should be noted that the random sequences produced by this method will become normalized, so that the mean value $\langle r_i \rangle$ is set to zero and its variance is kept fixed.

The phase transition which was analytically proved at the band center in Sec. III B [and Fig. 1(a)] is supported by the numerical calculations seen in Fig. 6(a). Figure 6(a) shows that $N\gamma$ increases with system size in the region below the critical Hurst exponent ($H_{cr}=1/2$) and remains fixed for $H>1/2$. This transition can be extended to other energies near the band center. A special energy near the band center $E=0.1$ is being considered. Figure 6(b) shows a similar behavior as if the energy were at the band center. As is clear from Fig. 6, away from the band center and for $H<1/2$, states are more localized than the band center energy. This localization property will be also seen in Fig. 7.

FIG. 6. (Color online) Numerical result of $N\gamma$ in terms of the Hurst exponent for $\sigma_{\ln(t)}=0.1$. (a) For $E=0$. This result confirms the localization-delocalization transition which is extracted by analytical calculations (Fig. 1). (b) For $E=0.1$. This transition can be expanded to the vicinity of the band center.

As was shown in Sec. III B, at $E=0$, the localization length has a power-law behavior with respect to the system size for all $H$. This exhibition of the power-law behavior is seen in Fig. 7(a). The localization-delocalization transition occurs at $H_{cr}=0.5$ when $\lambda \propto N$. Our analytical results on $\beta$ (exponent of $N$) versus $H$ at $E=0$ [Eq. (18)] were confirmed numerically as can be seen in Fig. 1(b)(points with error bar).

The localization-delocalization transition at $E=0.1t_0$ is again investigated as can be seen in Fig. 7(b) where for $H \geq 1/2$, the state shows a delocalization property. Furthermore, it is clear that for $H<1/2$ this state displays a dual behavior with respect to size. In fact, there is a critical system size below which there is a power-law behavior as $N^{1/2+H}$ (band center behavior with chiral symmetry). Above that size, the localization length becomes independent of size and one is in the localized regime. Therefore, as is shown in Fig. 6, the state at the band center is more delocalized than a state away from the band center. The critical system size which separates the chiral symmetry region from the localized regime increases with $H$. It tends to infinity in the critical Hurst exponent.

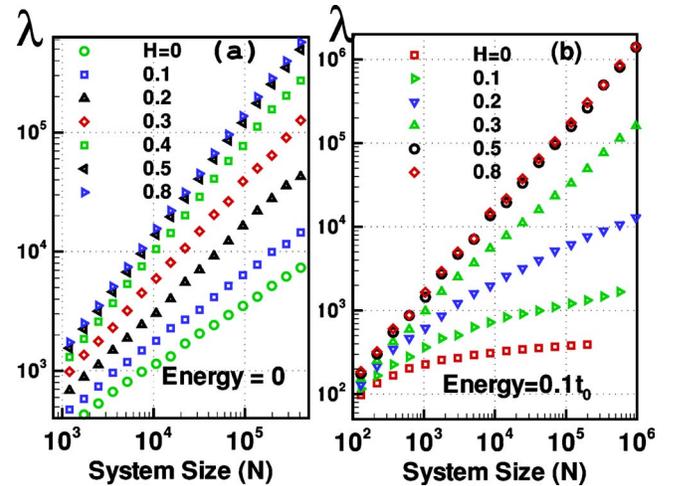

FIG. 7. (Color online) Localization length as a function of system size for different values of correlation exponent for (a) $E=0$ and (b) $E=0.1t_0$. Here $\sigma_{\ln(t)}$ is considered to be 0.1. The critical Hurst exponent ($H=1/2$) is the same as the band center one, and also it is equal to the critical exponent of the on-site disorder case. The localization length has a power-law behavior versus size at the band center.

In the localized regime, the exponent of energy $\eta$ is the same as the uncorrelated case, Eq. (21), and independent of $H$. The localization length versus $E$ is plotted in Fig. 3(b) for different values of $H(<1/2)$. Similar to the uncorrelated case and independently of $H$, the localization length for $H<1/2$ is proportional to $\sigma^{-1}$ (band center behavior) for $\sigma \ll \sigma_{cr}$ and has the form $\sigma^{-2}$ for $\sigma \gg \sigma_{cr}$ as can be seen in Fig. 5(b), where the critical variance is defined by using Eq. (23). Consequently, the following scaling law can be postulated for the localization length of long-range correlated disorder away from the band center:

$$\lambda_{\text{corr}}(E \neq 0) \propto \begin{cases} \frac{N}{\sigma_{\ln(t)}}, & H \geq \frac{1}{2}, \\ \frac{N^{1/2+H}}{\sigma_{\ln(t)}}, & N \ll N_{cr}, \quad H < \frac{1}{2}, \\ \frac{P(H)E^{-\eta}}{\sigma_{\ln(t)}^2}, & N \gg N_{cr}, \quad H < \frac{1}{2}, \end{cases} \quad (23)$$

where the critical length behaves as $[P(H)E^{-\eta}/\sigma_{\ln(t)}]^{1/(1/2+H)}$. The critical length and therefore $P(H)$ go to infinity when $H$ tends to the critical exponent. Hence, the function $P(H)$ is considered to have the form $(1/2-H)^{-\nu}$. The exponent value of $\nu(=6.5\pm1.0)$ is obtained by fitting $P(H)$ to $\lambda(H)$ in Fig. 7(b) for a fixed system size. The error is due to the slow saturation of the localization length versus size. The scaling laws for uncorrelated disorder can be obtained from the above equation by setting $H=0$.

## V. CONCLUSION

In conclusion, 1D localization properties of uncorrelated and correlated purely off-diagonal disorder have been studied analytically at the band center and numerically at other energy values near the band center. We derive an analytical expression for the localization length at the band center in terms of the pair correlation function. Long-range correlated disorder, which is normalized in variance, leads to a localization-delocalization transition at the critical Hurst exponent $H_{cr}=1/2$. This critical exponent is the same as for the on-site disorder case. At the band center, the localization length has a power-law behavior with respect to size for all $H$. In the thermodynamic limit, $\lambda$ is proportional to $N/\sigma_{\ln(t)}$ for $H\geq 1/2$ (delocalized) and $N^{1/2+H}/\sigma_{\ln(t)}$ for $H<1/2$ (anomalously localized), respectively. It is obvious that correlated disorder without any normalization process results in completely localized states for all $H$.

Away from the band center and for $H\geq1/2$, the system behaves as if the energy were at the band center. There is a dual property for $H<1/2$, however: for a length smaller than a critical size, $\lambda$ shows a power-law behavior similar to the band center case where the SPS theory breaks down, and for systems larger than this critical size, one is in the localized regime where the localization length varies with energy as $E^{-\eta}/\sigma_{\ln(t)}^2 (\eta \approx 0.18 \pm 0.03)$ independently of size.

## ACKNOWLEDGMENTS

We wish to acknowledge Professor M. Rahimi-Tabar and Dr. A. Bahraminasab for providing initial random data and useful discussions and Dr. A. T. Rezakhani for a critical reading of the manuscript.

## APPENDIX

The function **F** is a summation of random variables which have a Gaussian distribution. Equation (8) can be rewritten in the notation

$$\mathbf{F} = \sum_{i=1}^{N} U_i, \quad U_i = (-1)^{i+1}\ln\left(\frac{t_i}{t_0}\right). \quad (A1)$$

Wick's theorem can be applied to random variables with a Gaussian distribution. It can be proved that **F** has also a Gaussian distribution. Higher moments of **F** can be written as

$$\langle F^2 \rangle = \sum_{i,j} \langle U_i U_j \rangle,$$

$$\langle F^4 \rangle = \sum_{i,j,k,l} \langle U_i U_j U_k U_l \rangle. \quad (A2)$$

According to Wick's theorem, all higher moments are related to the variance of **F**:

$$\langle F^4 \rangle = \sum_{ijkl} (\langle U_i U_j \rangle \langle U_k U_l \rangle + \langle U_i U_l \rangle \langle U_j U_k \rangle + \langle U_i U_k \rangle \langle U_j U_l \rangle)$$

$$= 3\langle F^2 \rangle^2, \quad (A3)$$

where odd powers of **F** are equal to zero as $\langle F^n \rangle = 0$ for $n$ odd. The following expression implies that the function **F** has a Gaussian distribution:

$$\langle e^F \rangle = 1 + \frac{\langle F^2 \rangle}{2!} + \frac{\langle F^4 \rangle}{4!} + \cdots = e^{\langle F^2 \rangle/2}. \quad (A4)$$

Therefore, Eq. (11) still can be used for correlated disorder. In this equation, the LE is proportional to the variance of the **F** function ($\sigma_F^2$) which results in the calculation of $\langle F^2 \rangle$. By definition of the correlation function as $g(i-j) = \langle \ln(t_i/t_0)\ln(t_j/t_0) \rangle$, we can write down $\langle F^2 \rangle$ as the following:

$$\langle F^2 \rangle = Ng(0) - (2N-1)g(1) + 2(N-2)g(2)$$

$$- \cdots + 2g(N-1). \quad (A5)$$

This equation results in Eq. (14) in the text.